\documentclass{article}
\usepackage{amsmath}
\usepackage{hyperref}
\usepackage{graphicx}
\usepackage{xcolor}
\usepackage{multicol}
\usepackage{tikz}

\title{The Building Blocks of Consciousness}
\author{Robin W. Spencer\footnote{The author is a retired biochemist and data analyst.  Following degrees in physics (BA Williams College) and biochemistry (Ph.D. MIT), he worked at Syntex, Pfizer, and Imaginatik.  His current principal interests are research and software development in genetic genealogy. Contact spencerrw@alum.mit.edu, website \url{http://scaledinnovation.com} }}
\begin{document}
\sloppy   
\maketitle

\begin{abstract}Consciousness is presented not as a unified and uniquely human characteristic, but rather as an emergent property of several building blocks, most of which are demonstrably present in other species. Each block has its own rationale under natural selection and could have arisen independently, and the jumps between blocks — which culminate in consciousness — are small enough to be evolutionarily plausible. One underappreciated block involves unconscious engram playback and discrimination, and plays a major role in brain storage optimization. This function is present in birds and nearly all mammals and is recognized by its side-effect: dreams.
\end{abstract}

\section*{Introduction}
    In his book ``Science and the Soul", Richard Dawkins poses that the great outstanding problem for this century is how the human brain works, and specifically the nature of subjective consciousness$^{1}$. Others have said the same, though sometimes with as much discussion of philosophy as of science$^{2}$.  Any proposal about the elements and origin of consciousness must be consistent with observations and experiments with humans and other species, and must be overtly consistent with natural selection, preferably with a direct evolutionary narrative.
    In this paper I reject any premise that consciousness is irreducible, uniquely human, or ``special" in a philosophic sense.  Instead I suggest that there is a set of building blocks that create an evolutionary path for consciousness.  The literature is vast and certainly not all of what follows is original, though perhaps there are some new ideas to carry forward: 
    \begin{enumerate}
        \item Human consciousness should not be viewed as an irreducible, magical whole, but rather as an emergent property of a set of building blocks. Some but not all of these blocks are present in other species -- just as other species vary in the degree to which we'd say that they could be conscious.
        \item Each building block, perhaps with one exception, has the essential property that we can understand its independent selective advantage. This removes the need for ``one big jump" and provides path options for the evolution of consciousness.
        \item The blocks may have different prerequisites, either dependent on another block or dependent on some other property of a species. This means that some evolutionary paths could be parallel but others must be sequential. Several of the blocks require communication between species members.
        \item Some of the blocks involve beneficial communication between individuals of the same species; bees dance to point to food, sentinel horses warn the herd, apes teach each other.  A key jump is when such communication turns inward and becomes awareness.
        \item If a species has enough of the blocks, their synergy may generate its own selective advantage, which could be the force that drives to the property we call consciousness -- paraphrasing Hofstader, that point where ``the soul [becomes] greater than the hum of its parts$^{3}$.
        \item Crick and Mitchison$^{4}$ provide a clear, simple, data-supported function for dream sleep with a strong selective benefit. Importantly for this discussion, that function implies the properties of a building block that is critical for consciousness.
        \item The concepts here could be applied to machines: an experimenter could develop a set of analogous machine building blocks independently, then combine them under a common goal architecture and watch for emergent outputs.
    \end{enumerate}

    Since I have been pondering this topic for a long time$^{5}$, I am surely guilty of forgetting whom to acknowledge for some of these ideas. There is no intent to deceive or assert credit or precedence: my goal is simply to express the overall concept.
    
\section*{The Building Blocks}
    The blocks are in order of decreasing prevalence among species, and so also their likely time of appearance in evolution.    
    \begin{enumerate}
    \item \textbf{Intra-Species Communication.}
        A key tenet of this proposal is that consciousness arises from beneficial traits of social species, and so is not to be expected in asocial species (or isolated robots!). Of course signaling between individuals is ancient and not limited to vertebrates: bacteria and plants benefit from chemical signals without anything like human intent. Bees dance, ants leave chemical trails, insects and frogs fill the night with chirping. Whales sing, porpoises chatter, apes call, birds sing, all conveying information to each other. The point here is that communication is widespread and need not be conscious or intentional (in a human sense) to be useful and under strong selection pressure (find a mate, find food, repel or avoid threats).
        
    \item \textbf{Responsive Memory.}
        Birds, insects, and many animals play back migration programs twice a year. Seasonal changes in light or temperature trigger the sequence, at least for lead organisms: followers may use another set of rules such as swarm behavior. Likewise, the building of nests and dens and fleeing from the sounds and smells of predators are multi-step sequences triggered by external stimuli.

    \item \textbf{Learning.}
        Active learning is experimentally observable from Aplysia onward. Edward Thorndike's puzzle boxes showed that cats' learning might be random and gradual, but once learned, a sequence can be remembered and replayed. Skinner extended such experiments to complex chains of triggerable behaviors.~All of this demonstrates that many species have, and can acquire, brain-encoded scripts that we can observe played back in their behavior when elicited by an appropriate stimulus. What matters is that a trigger elicits a memory-fragment (engram), or chain of fragments, that plays out in behavior. The selection pressure for both instinctive and learned recall is obvious.
            
    \item \textbf{Unconscious Engram Playback Machine and Discriminator.}
        Recent optogenetic techniques have been used to trigger engrams in mice: a learned fear response (``freezing") can be replayed by external stimulation of a very small number (around ten) of specific hippocampal neurons$^{6}$.  The mouse does not see or control the electooptical stimulus, so its playback of the freezing behavior is involuntary.
        
        But more profound than these reductionist experiments is the occurrence of dream sleep.  Crick and Mitchison$^{4}$ make the compelling argument that dream sleep has a sensible, useful purpose:  dreams are simply the remembered residue of a process that cleans up our brain-storage system by deemphasizing nonsense connections, analogous to the defragmentation of a computer disk.  With such a direct, mechanical purpose, dream sleep can be seen to be under strong selection pressure, consistent with its presence in birds and all marsupial and placental mammals.  As a side-effect it is  philosophically refreshing that Crick and Mitchison wield Occam's Razor powerfully enough to slice through millennia of balderdash. 
            
        The function of REM sleep implies that four very important elements are present:  first, that blocks \#2 and \#3 are present:  there is a selection of engrams, some new and some old, and second, that there is an internal mechanism that can trigger the playback of an engram.  It matters that the trigger is internal, not driven by the senses.  Third, there is a mechanism to string engrams together into a sequence (as evidenced by dreams' 20-30 minute length, remembered narrative complexity, and weird jumble of topics), and fourth -- very important -- there is a mechanism to judge the relevance or importance of engrams to each other.  This latter discriminator is central to Crick and Mitchison's hypothesis, since the active de-weighting of nonsense connections is integral to the more efficient packaging that is the selection pressure behind the evolution of dream sleep.  So dreaming species already have a way to discriminate between ``sensible" connections of engrams that are relevant when awake$^{7}$, versus nonsense (or even antisense) connections which when purged do little harm, clear up brain-space, and of course are sometimes remembered when we wake as dreams.
             
        The beautiful thing here is that all of this is under straightforward selection pressure -- more brain capability at a given size.  Because a mouse dreams does not mean that it is conscious in the human sense, merely that it is using an ancient brain-efficiency mechanism while asleep$^{8}$.             
             
    \item \textbf{Sentinel and Observer.}
        My example for this block is the transfer of awareness in a clan/flock/group.  For example, a wolf or horse on the periphery of the pack may sense something outside, and quite unconsciously stand more stiffly and lift its ears.  Another pack-mate, while not sensing any threat, may nonetheless see the sentinel and itself come to an alert state.  This is certainly a useful trait and would be under positive selection, and since the benefit goes to the internal observer without having to invoke group selection$^{1}$.
         
        This block is more specific and arguably more advanced than swarm or stampede behavior in insects, birds, prey mammals, and fish -- those take some critical mass of pack-mates in motion to get others to join the mass movement, and the response action is the same as the observed action (close-pack flying, swimming, or running).   What sets this block apart from \#1 on general intra-species communication?  This block has a specificity of individuals, action, and timing.  It emphasizes observable triggers and behaviors within a pack.  The communication is one-way:  the sentinel would come to alert if it were alone; its fight-or-flight reflex is beneficial regardless of any of its pack-mates watching.  The observer senses the sentinel, but not the other way around.
         
        This block is related to the case in which one dog will adjust its actions depending on the actions of another dog, and not for mortal stakes but in play$^{9}$.  
        
    \item \textbf{Self Recognition.}
        This block is defined by the ``forehead dot and mirror" experiment, in which a subject is allowed to see itself in a mirror for a period of time, secretly labeled with a dot of paint, and again presented with the mirror.  While many animals show social reactions to their mirror image (threat or avoidance), a few animals will scratch or rub the dot, demonstrating a sense of self.  Chimpanzees, orangutans, bonobos, dolphins, elephants, magpies, cleaner wrasse fish, and humans after age 18 months pass the test, and many others fail (including macaques, monkeys, other birds and fish)$^{10}$.
        
        While this block is easily tested experimentally, it is the only block for which there is no satisfying and direct hypothesis to suggest its benefit and therefore its selection pressure.  However the next block -- Theory of Mind -- has a clear selection narrative, and the sequence of appearance of these two blocks in human children and their narrowing prevalence in other species suggests that self-recognition may be a necessary step toward Theory of Mind.
        
    \item \textbf{Theory of Mind.}
        An individual with a theory of mind will attribute mental states to itself and others.  When chimpanzees are presented with a video of a human struggling with a complex puzzle followed by a set of photographs of puzzle solutions, they will choose the correct answer.  Ravens and bonobos may have related abilities$^{11}$.  This requires a higher level of abstraction and learning than the sentinel-observer behavior.  Most children develop this skill from ages 3 to 5, and it can be deficient in the autistic or schizophrenic.  Parents' ability to impute thoughts and emotions in their children is clearly beneficial in child raising and supplies selection pressure for this block$^{12}$.

    \item \textbf{Emotive Consciousness.}
        This is the whole package characteristic of healthy human adults, with sensory awareness, social skills including empathy, creative problem solving, and the ability to originate and continue a narrative train of thought.  Our dominance as a species attests to its benefits in natural selection.
        
    \item \textbf{Flat Consciousness.}
        To allow a range of properties for human-level consciousness, I've split it into two blocks.  Our principal concern is with Emotive Consciousness above, but certainly there is a range.  Some people have reduced or largely absent theory of mind, but they are conscious in that they respond to their senses, listen and talk. They are not zombies (a loaded term) or robots.  I presume -- but do not know, and have not found literature -- that such people dream, pass the mirror self-recognition test, and can solve many types of problems. This state is not so much a building block in a sequence, but a state that may be revealed by disease or trauma; in isolation it may not be supported by selection pressure, but revealed by a deficit from a state that has selective advantage, and for this reason is shown as a deficit state from emotive consciousness in the diagram below.   It is a reminder than full emotive consciousness is fragile.
    \end{enumerate}

\section*{Connecting the Blocks}
    Figure 1 suggests a sequence for the appearance of the blocks over evolutionary time.  The color scheme shows that some species carry forward while others stop at a given point and do not acquire the next block (Why?  There is no why with evolution -- it is simply good enough).

    Blocks may also differ in degree, across species and over the lifetime of the individual.  Specifically the responsive memory block will grow with experience.  Where there are quantitative changes there can also be thresholds, so, for example, it is possible that a human child may have the genetic wiring for Theory of Mind, but until it acquires enough experiences, it simply does not have a rich enough library of memory fragments to show the behaviors characteristic of that block.  You can't reason by analogy if you don't have enough material for comparison.
        
    \begin{figure}
    \centering
    \includegraphics[width=13cm]{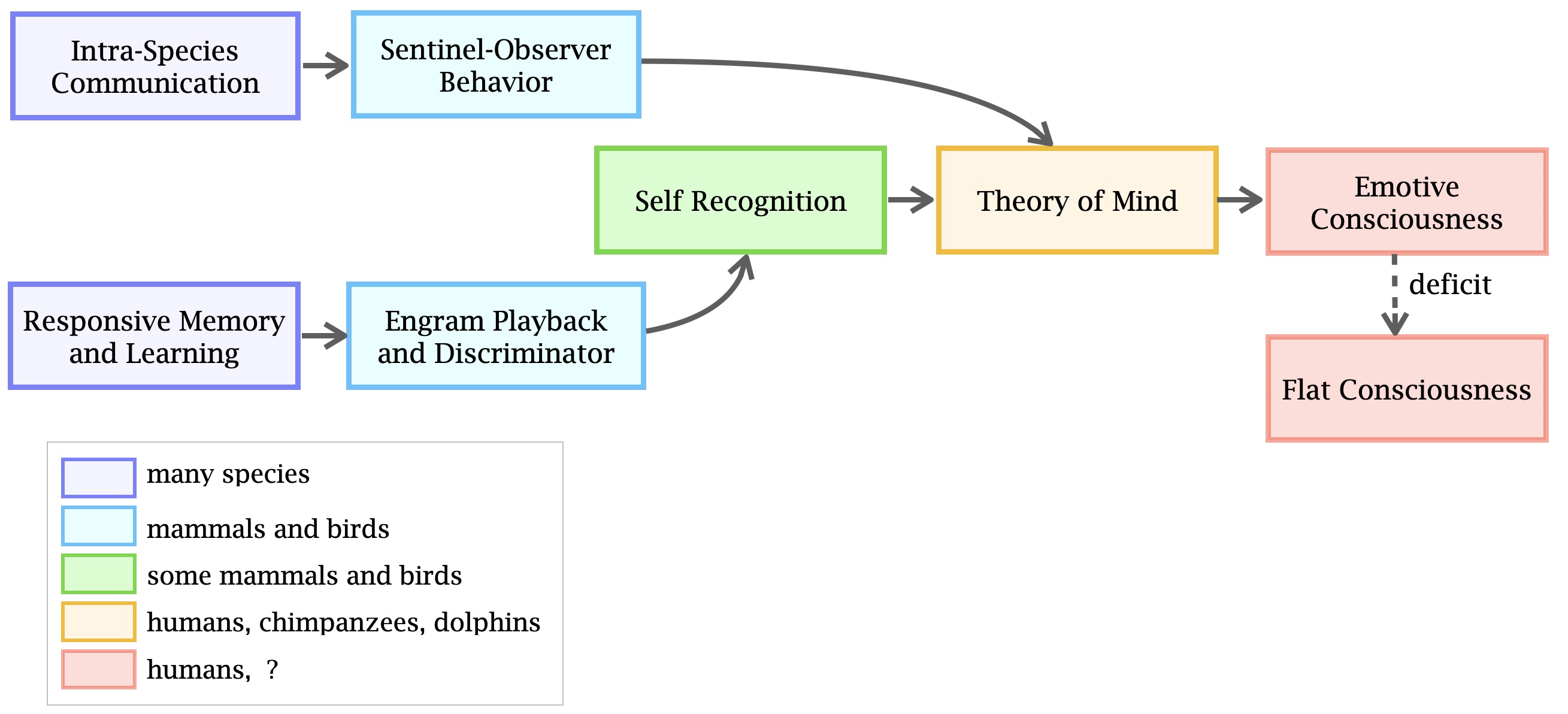}
    \caption{The building blocks of consciousness}
    \end{figure}

    Others have described something like the upper pathway.  My contribution is the emphasis on \textit{communication between individuals}, which provides an ancient anchor and selection pressure along the way (and this does not preclude other paths contributing to Theory of Mind). My point is simply to show that at least one set of observable behaviors (related to beneficial communications between group-mates) provides a prerequisite for consciousness:  for example empathy, considered a very high level property, is all about the observation of subtle signals from a clan-mate (facial expression, etc), inference of their mental state, and reaction to that state.
    
    Also as noted above I do not have a convincing narrative for selection pressure to arrive at self awareness as defined by the dot-and-mirror experiment.  That is worth some effort, but for the moment it is sufficient that this block is tightly defined by a simple experiment and observed in multiple species and not others.
    
    One of the key suggestions here is that it is a small jump$^{13}$ from having both the Sentinel-Observer and Self Awareness blocks to the Theory of Mind block.  For example: I hear a noise in the bushes and notice that I'm sitting up and my head is craned forward (self awareness).  I remember that I've seen this posture in a clan-mate on sentinel duty, and it put me on alert (sentinel-observer).  Therefore my current state is ``alert", and I queue up relevant behavioral scripts (fight or flight).  In this context, Theory of Mind is a generalized, useful combination of its two preceding blocks.
    
    The other key element of this article is the lower pathway and specifically the implications for consciousness of the memory-optimization engine that operates while we sleep -- the engram playback and discriminator block. Recapitulating, the existence of this engine implies that the organism
    \begin{itemize}
    \item has a retrievable memory storage system
    \item can chain memory fragments (engrams) together
    \item can unconsciously distinguish between related and unrelated fragments
    \item can, without sensory input, trigger such a chain of engrams which may persist for several minutes
    \end{itemize}
    Work since Crick and Mitchison (for example a role for non-REM sleep in the consolidation of short-term  memory -- a positive discriminator$^{14}$) expands their theory and remains consistent with these properties.

\section*{The Properties of the Memory Chunks}

    The optogenetic fear-recall experiments$^{6}$, our remembrance of dreams, and observations of scripted animal behavior all suggest that the relevant ``chunk size" of memory is about what can be played back in a specific action and with content comparable to a paragraph of prose (more than a single character, less than a novel).  For example, a dream about flying naked over Hyde Park might be assembled from at least three such chunks:  flying, being naked, and what Hyde Park looks like.  Likewise an animal's migration -- a large script -- would be composed of many unit scripts, chained together in combinations that can be biased by sensory input.  Note that this chunk size is very different from the single-symbol chunk of Searle's Chinese Room argument.
        
    It simplifies the model to suppose also that memory chunks contain not only content but also links to other chunks or to actions.  Consider books in a library: as they stand, the books are independent.  But what if each book, at the end, also contained a list of recommended sequels.  You might follow a path from colonial architecture to interior design to woodworking -- or with slight nudges, a different path from the same starting point.  Fragments combine to become stories, and engrams become thoughts (awake) or dreams (in REM sleep).
            
    Wikipedia provides a metaphor: each page of Wikipedia is a self-contained chunk of information.  But each page is also sprinkled with blue links, by which the reader may branch off to explore something related, perhaps a definition, an example, or an in-depth elaboration of the linked phrase.
        
    Some philosophers base their thought-experiments on computers that deal with individual symbols as their chunk-size, which when isolated are devoid of meaning or capability.  But that's much too small a chunk, for which software development provides a metaphor:  for a long time now, programmers have used object-oriented programming (OOP) to express their ideas.  In OOP, an object may contains both data (nouns) and methods (verbs).  A construct called \textit{myTable} might contain names, addresses, and telephone numbers, and also contain the ability to act on those data with commands like \textit{myTable.sort} and \textit{myTable.print}.  The programmer understands the high-level chunk written in the high-level language, but cannot understand the final compiled code -- and does not need to, since it is the compiler's complex but entirely mechanical task to do the one-way translation.  The relevant chunks for thought-experiments about computers are not isolated 64-bit symbols, but rather analogously rich objects, just as the relevant chunk-size for our wet-ware is the memory engram or simple action, not the synapse or action potential.
   
    This hypothesis is unchanged if instead some chunks are content-type (``Hyde Park") while others are action-type (``How to alphabetize a list").  We have established a chaining mechanism and a relevance evaluator, so this is equivalent to having instructions within the chunk.  Programming before OOP (``procedural programming") took this form, in which data and methods were separate, but functionally attachable, entities.
        
    Having memory-chunks which are both chainable and actionable removes the need for a Cartesian homunculus.

\section*{Internal Narratives}
    Dream sleep is evidence that birds and most mammals can assemble and filter memory fragments into extended mental narratives.  It is a small jump to suggest that this capability plays a significant role in consciousness and thought -- small enough to attempt a list of properties of human-style consciousness that are still unaccounted for:
       \begin{enumerate}
        \item a connection between sensory input and the triggering of an internal narrative
        \item given one fragment, find and invoke another that is related
        \item a mechanism to enable the narrative to elicit behavior -- as subtle as a raised eyebrow or rotated ear, or as complex as a spoken language 
        \item a mechanism to enable the narrative \textit{not} to elicit behavior -- such that the narrative plays out privately in the awake individual 
        \item a ``goal engine" that moves the narrative in a certain direction, and can end it
        \end{enumerate}
    
    Property (1) may already exist as part of earlier blocks; a conditioned stimulus will evoke a learned response in any animal.  If I say ``go" and point, my dog will race forward, stop about ten yards out, come to alert, and go into a search pattern: a combination of an innate script (she's a retriever) with learned behavior.  Property (2) is likewise not very difficult:  if the dream-engine can distinguish related from random engrams, then having this capability when awake is a small jump.  Nor is (3) hard to see:  dreams elicit eye movement and leg twitches, and many of the memory scripts mentioned above are connected to behavior.  When an antelope smells a predator on the wind, does it ``know" that there's a lion nearby? That's a moot and human-centric point of view, since the antelope's flight is what matters.  All we need to invoke is that the olfactory system's pattern finds a match with the encoded memory for lion-smell, which is chained (by long selection pressure) to a flight script.  Property (4) -- playback without behavior -- already exists for dreams, in the sense that most of the body is frozen while the dream-script plays.  Given the advantage of the sentinel-observer block, you can make the argument that the ability to suppress any external evidence of a running memory-script could also be a selectable advantage:  the equivalent of a good poker face.  So (4) is not problematic.
    
    Property (5) -- the presence of a goal-engine -- is the most interesting.  The dream-time memory compressor either operates without a goal-engine, or perhaps more simply just chains together engrams at random and down-weights those connections to simplify the overall network -- so we don't get clues about (5) from the dream-engine.  Animal migrations seem incredibly purposeful and driven, but their occurrence in insects suggests a sequencing mechanism far earlier and farther from human-type consciousness than we seek here.  An interesting argument for how-hard-can-it-be comes from computer science:  in recent years chess, go, and poker have been conquered by machines, notably not by human-scripted rules or exposure to master-level games (as were the original chess programs), but rather by a simple set of rules, a simple definition of success, and then practice, practice, practice over millions of games, played by the software against itself.  It is the simplicity and success of this approach which suggests that a goal-engine is not a terribly difficult element to propose, even though the meat-ware version would be implemented by natural selection in a manner we do not know.
    
    One possibility: imagine a simple goal (``get food"), a rich storehouse of memory-fragments, and a relevance-comparator (which we use to select what \textit{not} to chain together during the dream-cleanup process).  If the goal statement has higher overlap with one engram (prey along the riverbank) than another (a barren mountainside), then just following the stronger chain, link by link, is more likely to move toward the goal.  Of course this is speculation -- but this simple follow-the-chain concept is sufficient to build practical software that finds connections in hypotheses about disease$^{15}$.
    
\section*{Human Limitations and Machine Incursions}
    There is an interesting segment in the ``four horsemen" discussion with Richard Dawkins, Daniel Dennett, Christopher Hitchens, and Sam Harris$^{16}$ in which these highly educated and conscious human beings agree that none of them really comprehends quantum mechanics, but nonetheless are comfortable following its rules to achieve accurate results.  In a related vein, large software systems are arguably the most complex constructs ever built by humans$^{17}$, and are built in modules that encapsulate complexity and expose minimalist connections under very tight rules:  otherwise they would be too large for a single person to comprehend, or have too many internal dependencies to be built by teams.  These are practical consequences related to the observation by Herbert Simon and Malcolm Gladwell, namely that it takes ten years or ten thousand hours of effort to achieve world-class expertise in any given endeavor$^{18}$.

    Thus our stone-age-evolved brains are good enough for the task of survival in the savannah, but not necessarily much more$^{16}$.  Human brain-power and consciousness are not the the end-all-and-be-all but just the point at which we find ourselves.  We know that our capabilities can be readily degraded by trauma or disease, and at some point would drop below any reasonable definition of consciousness.
    
    The flip side of degradation of consciousness would be its advancement.  The Gladwell/Simon number suggests that the limit has been reached:  any capability that required 3-10 fold more study and practice would not have enough time for creative utilization within a human lifetime.  Even if culture does not overwhelm natural selection, we do not have the patience to wait 100,000 years for the next version of \textit{Homo sapiens}.  And so, for many of the kinds of experiments needed to address hypotheses about consciousness, we must turn to machines and software.

    First example: imagine Richard Dawkins carrying out a calculation in quantum mechanics.  Second example: when I write software, I usually adopt someone else's code to do (for example) some complex procedure in matrix algebra.  I check for the accepted answer with a couple of test datasets and then happily use the code without concern.  Neither of us has an intuition for the innards of the method (i.e. could not explain it) but we get the right answers.  So -- what's the difference between us and a computer that plays chess or go or poker?  Not that much, actually.  This train of thought quickly leads to the Turing test -- which, if taken seriously, stands up as a valid test of consciousness$^{2}$, albeit a very human-centric kind of test$^{19}$.
    
    Chess, go, and poker playing software are examples of ``push" systems deliberately built to test a milestone task.  We also have increasing evidence of ``pull", in which humans want what technology can offer and the marketplace responds. Regardless of financial and privacy costs, millions of people invite Siri and Alexa into their lives, and within their sphere (especially travel and entertainment advice), they have become preferred sources for basic factual information.  Likewise, personal robots are making inroads into elderly care, and their mechanical responses can hold up one side of a short conversation enough to gladden the lonely.   Are these devices conscious?  Of course not -- the point here is simply that we humans are increasingly interacting with them as if they were, and to that extent they are moving closer to passing a Turing test.  Curiously, high-quality voice recognition and synthesis, though entirely mechanical and unmysterious, provide a major jump in our acceptance, such that these capabilities become building blocks of ``good enough machine intelligence." $^{20}$
    
\section*{Conclusions}
    It would be gratifying if anyone has, or would, take some of the ideas here into machine intelligence research.  This could entail building enough systems to form a little clan, and giving each a mechanism for inter-entity communication.  At least one of the communication modes should be accessible to the entity as well as its peers, so that it can ``sense when its own ears are pricked up."  A common set of goals would be useful, perhaps just survival in some virtual ecosystem where persistence is not assured and decisions have consequences.  A memory of diverse fragments/engrams/memes, as large and as cross-linked as Wikipedia, would be necessary$^{21}$.  Analogs of the building blocks presented here, as well as others, would be interesting to add to some entities and not others -- to have different ``species" with different capabilities in the system, perhaps in the sense of predator and prey.  Learning and memory are a given.
    
    I hope that the reader has found some ideas here that are interesting and stimulating.  Much of the literature of consciousness is, to my eyes, embarrassingly human-centric, predetermined in its judgment (``Of course we're special!"), and tilted too far toward philosophy and provocation (seeking \textit{why}) and not enough toward natural selection and problem-solving (seeking \textit{how}).
    
\section*{Footnotes and References}

\begin{enumerate}
    \item Dawkins, Richard, 2017, ``Science and the Soul", Transworld Publishers, London, p 95
    \item Dennett, Daniel, 1991, ``Consciousness Explained", Little, Brown, Boston
    \item Hofstader, Douglas and Dennett, Daniel, 1981, ``The Mind's I", Basic Books, New York, p 191
    \item Crick, Francis, and Mitchison, Graeme, 1983, ``The function of dream sleep", Nature 304: 111-114  \url{https://profiles.nlm.nih.gov/ps/access/SCBCDK.pdf}
    \item ever since Douglas Hofstader, 1979, ``Gödel, Escher, Bach: an Eternal Golden Braid", Vintage Books, New York, p 191
    \item Liu, X., Ramirez, S., Pang, P. T., Puryear, C. B., Govindarajan, A., Deisseroth, K., \& Tonegawa, S., 2012, ``Optogenetic stimulation of a hippocampal engram activates fear memory recall", Nature, 484(7394), 381.
    \item A useful pairing might be ``I remember hearing growling in those woods" with ``there's a climbing tree near my den."
    \item Siegel, J. M., Manger, P. R., Nienhuis, R., Fahringer, H. M., \& Pettigrew, J. D., 1996, ``The echidna Tachyglossus aculeatus combines REM and non-REM aspects in a single sleep state: implications for the evolution of sleep", Journal of Neuroscience, 16(10), 3500-3506.  This paper finds that the spiny anteater does not have comparable REM sleep to placental mammals -- which is why Crick and Mitchison suggest it has such an anomalously large brain (it's inefficient).  This puts the appearance of REM sleep in mammals as early as the Jurassic about 170 million years ago.
     \item Horowitz, Alexandra, 2009, ``Attention to attention in domestic dog (Canis familiaris) dyadic play." Animal Cognition 12.1: 107-118  \url{https://psychology.barnard.edu/sites/default/files/attentiontoattention.pdf}
     \item Gallup, Gordon, 1970,``Chimpanzees: self-recognition." Science 167: 86-87  \url{http://radicalanthropologygroup.org/sites/default/files/pdf/class_text_023.pdf}, also \url{https://en.wikipedia.org/wiki/Mirror_test}
     \item Premack, David, and Guy Woodruff, 1978, ``Does the chimpanzee have a theory of mind?." Behavioral and brain sciences 1.4: 515-526  \url{https://is.muni.cz/el/1423/podzim2012/PSY277/um/36127513/PREMACK_WOODRUFF_cimpanzees.pdf}  \url{https://en.wikipedia.org/wiki/Theory_of_mind}
     \item Many other beneficial narratives are easily imagined, for example, ``I won't steal food from that bigger tribe-mate.  He'd get mad and clobber me."
     \item By small jump, I mean what Dawkins[1] calls a ``stretch DC-8" jump as opposed to a ``Boeing 747" jump.  The former is readily accommodated by natural selection, while the latter appears to require a designer-god.
     \item Rasch, Björn and Born, Jan, 2013, ``About Sleep's Role in Memory", Physiol Rev. 93(2): 681-766  \url{https://www.ncbi.nlm.nih.gov/pmc/articles/PMC3768102/}
     \item Smalheiser, Neil R., and Swanson, Don. ``Using ARROWSMITH: a computer-assisted approach to formulating and assessing scientific hypotheses." Computer methods and programs in biomedicine 57.3 (1998): 149-153 
     \item The Four Horsemen - Hitchens, Dawkins, Dennet, Harris, 2007, video \url{https://www.youtube.com/watch?v=n7IHU28aR2E} .  Note that my argument is not about whether quantum mechanics is correct.  Quantum mechanics (like relativity) has experimentally demonstrated predictive power which puts almost at the level of mathematical proof:  we do not have to understand it (all) personally to believe it to be valid.  The difference from religion is of course the open, written, falsifiable-by-evidence process of math or science.  If we did not accept these shortcuts (trust in the work of others, acceptance of proof), we'd all be Flat Earthers: crippled by accepting only that which can personally observed with unaided senses.
     \item Brooks, Frederick P., 1995, ``The mythical man-month (anniversary ed.)", Addison-Wesley, Boston
     \item Gladwell, Malcolm, 2008, ``Outliers: The story of success." Hachette UK, and Simon, Herbert A., 2019,  ``The sciences of the artificial" MIT press
     \item Consider whether a human could pass a dolphin's Turing test, which wouldn't have a sealed room and a teletype machine, but would instead derive from 20 million years of natural selection in a three dimensional environment and with ultrasonic senses that carry for miles.
     \item Content plus voice communication = utility + convenience.  While we could use a desktop computer and the internet to find out local movie times 25 years ago, smartphones and always-on voice recognition devices drop the everyday usage effort to zero.  At some point, the ubiquity, utility, and convenience of systems can be seen as a kind of large-scale Turing test.  And that's without adding autonomous planes, ships, trucks, trains, and cars into the discussion.
     \item A vast and diverse memory is integral to a serious Turing test.  Dennett alludes to this in his refutation of Searle's Chinese Room argument$^{2}$.  The computer in the room doesn't just shuffle Chinese characters around -- that's a human-imposed crippling limitation.  Play fair and give the computer a Wikipedia-sized memory, so when the room is asked ``do you feel that you know Chinese?", it can bring up a long discourse on what it means to feel, much like a human pedant.
     
\end{enumerate}
 
\end{document}